# Machine-learning interatomic potential for molecular dynamics simulation of ferroelectric KNbO$_3$ perovskite

Hao-Cheng Thong,[1,2] XiaoYang Wang,[3,*] Han Wang,[3] Linfeng Zhang,[4] Ke Wang[1,*], and Ben Xu[2,*]

**AFFILIATIONS**

[1] State Key Laboratory of New Ceramics and Fine Processing, School of Materials Science and Engineering, Tsinghua University, Beijing 100084, PR China.

[2] Graduate School, China Academy of Engineering Physics, Beijing 100193, PR China.

[3] Laboratory of Computational Physics, Institute of Applied Physics and Computational Mathematics, Huayuan Road 6, Beijing 100088, PR China.

[4] Beijing Institute of Big Data Research, Beijing 100871, PR China.

*Author to whom correspondence should be addressed: xiaoyanglanl@gmail.com (X. Wang), wang-ke@tsinghua.edu.cn (K. Wang), bxu@gscaep.ac.cn (B. Xu)

**ABSTRACT**

Ferroelectric perovskites have been ubiquitously applied in piezoelectric devices for decades, among which, eco-friendly lead-free (K,Na)NbO$_3$-based materials have been recently demonstrated to be an excellent candidate for sustainable development. Molecular dynamics is a versatile theoretical calculation approach for the investigation of the dynamical properties of ferroelectric perovskites. However, molecular dynamics simulation of ferroelectric perovskites has been limited to simple systems, since the conventional construction of interatomic potential is rather difficult and inefficient. In the present study, we construct a machine-learning interatomic potential of KNbO$_3$ (as a representative system of (K,Na)NbO$_3$) by using a deep neural network model. Including first-principles calculation data into the training dataset ensures the quantum-mechanics accuracy of the interatomic potential. The molecular dynamics based on machine-learning interatomic potential shows good agreement with the first-principles calculations, which can accurately predict multiple fundamental properties, *e.g.*, atomic force, energy, elastic properties, and phonon dispersion. In addition, the interatomic potential exhibits satisfactory performance in the simulation of domain wall and temperature-dependent phase transition. The construction of interatomic potential based on machine learning could potentially be transferred to other ferroelectric perovskites and consequently benefits the theoretical study of ferroelectrics.

**KEYWORDS**: Molecular Dynamics, Machine Learning, Interatomic Potential, Ferroelectric, Perovskite

## I. INTRODUCTION

Since the discovery of ferroelectricity by Valasek in 1920, the development of ferroelectric materials has been an important branch of scientific research.[1] Over the past century, numerous ferroelectric materials were discovered and have been ubiquitously used in applications, *e.g.*, electromechanical transducers[2], memory devices[3], and energy storage devices[4]. Among various ferroelectrics, the family of perovskites has been a magnificent choice for these applications, as they usually exhibit a stronger spontaneous polarization and enhanced dielectric properties, *e.g.*, application of $BaTiO_3$ in capacitors.[5]

High-precision and responsive piezoelectric devices have been an essential component of the development of modern technology. Ferroelectric perovskites are heavily applied in piezoelectric devices, where $Pb(Zr,Ti)O_3$ (PZT) ceramic is the most famous one.[6] Domination of commercial PZT ceramic for almost 50 years has been recently challenged by the environmental concern over the contamination of the hazardous lead element.[7] In response to the concern, the development of lead-free substitutes has grown rapidly. Multiple groups of lead-free ferroelectric perovskites have been demonstrated to be excellent candidates, *e.g.*, $(K,Na)NbO_3$-based, $BaTiO_3$-based, $(Bi,Na)TiO_3$-based, and $BiFeO_3$-based ceramics.[8] Among these perovskites, $(K,Na)NbO_3$-based ceramics possess the most similar properties to PZT ceramics, including piezoelectric coefficient and Curie temperature, which is thus favorable for practical replacement in many scenarios.[9]

In parallel with experimental research, theoretical calculation has been an important strategy for investigating ferroelectrics. With the rapid advances of high-throughput computing and extensive construction of infrastructure global, theoretical calculation has become increasingly accessible to scientists over the past 20 years. While first-principles density functional theory (DFT) calculation provides microscopic insight into the electronic structure of ferroelectrics,[10] molecular dynamics (MD) simulation is advantageous for investigating the macroscopic properties of ferroelectrics, *e.g.*, dielectric response,[11] electromechanical response,[12] phase transition,[13] nucleation and growth of domain walls[14].

Even though, performing a realistic MD simulation is extremely difficult, for which an accurate interatomic potential (also frequently known as potential energy surface or force field) is a prerequisite.[10] Approximation from DFT calculations has been extensively utilized to obtain the interatomic potential with an accuracy of quantum-mechanics level. Numerous interatomic potentials were developed in the past, including the shell model,[15] bond-valence model,[16] and effective-Hamiltonian model.[17] These models have been successfully used to describe the lattice dynamics in various ferroelectric perovskites. However, since numerical fitting for these models is a complicated task, the construction of interatomic potentials is often limited to a simple system. Without proper mathematical treatment and the aids of automated toolkits, the construction of the interatomic potentials for a comprehensive system that consists of multiple phases (or even an alloy system), would be technically challenging.

Machine learning (ML), or more specifically deep learning, has been recently introduced for the construction of interatomic potential.[18-20] As the backbone of deep learning, a neural network is capable of extracting microstructural features from the atomic configuration of a many-body molecular system and establishing an approximated relationship between the microstructural features and fundamental properties (*e.g.*, interatomic force and energy) effortlessly.[21,22] The ML interatomic potential has been demonstrated to be a splendid model for the MD simulation of many different ferroelectrics.[23,24] ML is favorable for the construction of interatomic potential as it does not require a prior understanding of the system, thus it can be applied to various systems. However, the lack of physical meaning of the parameters in the model is sometimes considered to be a shortcoming, from some physicists' perspective.

In this work, the interatomic potential of ferroelectric $KNbO_3$ perovskite was developed via deep learning, as a representative system for lead-free $(K,Na)NbO_3$. The DFT calculation can be well reproduced by the MD simulation based on the ML interatomic potential, including the prediction of atomic force, energy, elastic properties, and phonon dispersion. On this basis, the MD simulations of the domain wall configuration and the intrinsic domain-wall motion were performed, without feeding any relevant configuration to the training dataset. Additionally, simulation of the temperature-dependent phase transition was attempted and compared with the experimental data. The MD simulation result of the ferroelectric $KNbO_3$ perovskite was satisfactory. Possibly, it can also be extended to other ferroelectric perovskite systems. MD simulation with improving accuracy can help deepen the understanding of the dynamics and physical properties of ferroelectrics from the theoretical calculation perspective.

## II. COMPUTATIONAL METHODS

The ML interatomic potential is constructed by using open-source software, DP-GEN.[25] DP-GEN is a concurrent machine learning platform that implements the "on-the-fly" learning procedure based on the DeePMD-kit package,[26] namely, the training of interatomic potential models and the generation of training data occur simultaneously. The design of the concurrent learning platform minimizes human intervention during the construction of the interatomic potential over a wide range of thermodynamic conditions, which can highly increase the work efficiency and lower the computational cost. One may refer to the original papers for technical details of DP-GEN.[25] Following the nomenclature previously used, the ML interatomic potential constructed by using DP-GEN is herein named deep potential (DP).[22] Below, we will briefly summarize several key parameters for the setup in DP-GEN.

Initiation of DP-GEN usually requires feeding of a few DFT calculations of a system at around the ground-energy state, for which 20-atoms cubic, tetragonal, orthorhombic (or monoclinic), and rhombohedral $KNbO_3$ configurations were provided. For the DFT calculation, an energy cutoff of 900 eV and a K-point mesh with a spacing of 0.12 were selected. DFT calculation within the generalized gradient approximation (GGA) with the Perdew-Burke-Ernzerhofer exchange-correlation function was performed by using Vienna Ab initio Simulation Package (VASP). The convergences of energy and force were set at the criteria of $10^{-6}$ eV and 0.005 eV/Å, respectively.

3 major procedures, *i.e.*, "training", "exploration", and "labeling" are included in DP-GEN iterative scheme. For "training", a descriptor of DeePot-SE[22] (consists of radial and angular information of atomic configuration) was adopted, with a cutoff radius for neighbor searching within 9 Å and smoothing cutoff from 1.5 Å. A 3-layers embedding net (20, 40, 60) and a 3-layers fitting net (120, 120, 120) were chosen. The "exploration" was performed at various temperatures (from 0K to 800K) and pressures (from 1 kbar to 10000 kbar). "Labeling", *i.e.*, generate DFT data corresponding to the poorly simulated configurations of MD trajectories during "exploration", was performed based on the model deviation of force within 0.12 to 0.30 eV/Å. Consecutive addition of necessary "labeled" data into the dataset for training during iterations is the key to the construction of interatomic potential.

## III. RESULTS AND DISCUSSION

### A. Examination of DP model

The experimentally known temperature-dependent phase transition of $KNbO_3$[27] is illustrated in Figure 1a. Above Curie temperature ~700 K, $KNbO_3$ is a paraelectric cubic phase (*Pm-3m*). Upon cooling down from Curie temperature, it goes through successive phase transitions, in a sequence as follows: tetragonal (*P4mm*), orthorhombic (*Bmm2*), and rhombohedral (*R3m*) ferroelectric phases. The spontaneous polarization directions of these ferroelectric phases are parallel to $<001>_{pc}$, $<101>_{pc}$, and $<111>_{pc}$ of the parent cubic phase, respectively. To obtain a comprehensive interatomic potential of the $KNbO_3$ system, these individual phases were evenly treated as a collective in DP-GEN. Roughly 10000 configuration data were generated in total during the exploration at different thermodynamic conditions, which should supposedly cover most of the configurations in $KNbO_3$ mentioned above.

Examination of the DP model is shown in Figure 1(b)-(c). The prediction of the DP model on atomic force and energy of all training data shows good agreement with the DFT calculation, where the error of energy and force are at the order of $10^{-3}$~$10^{-4}$ eV/atom and ~$10^{-2}$ eV/Å. The prediction error of the DP model is slightly larger than that of a usual DFT calculation which originates from functional approximation, energy break condition for electronic self-consistent loop, *etc.*, but it manages to provide reliable calculation on the properties of interest, which will be further demonstrated in the following paragraphs.

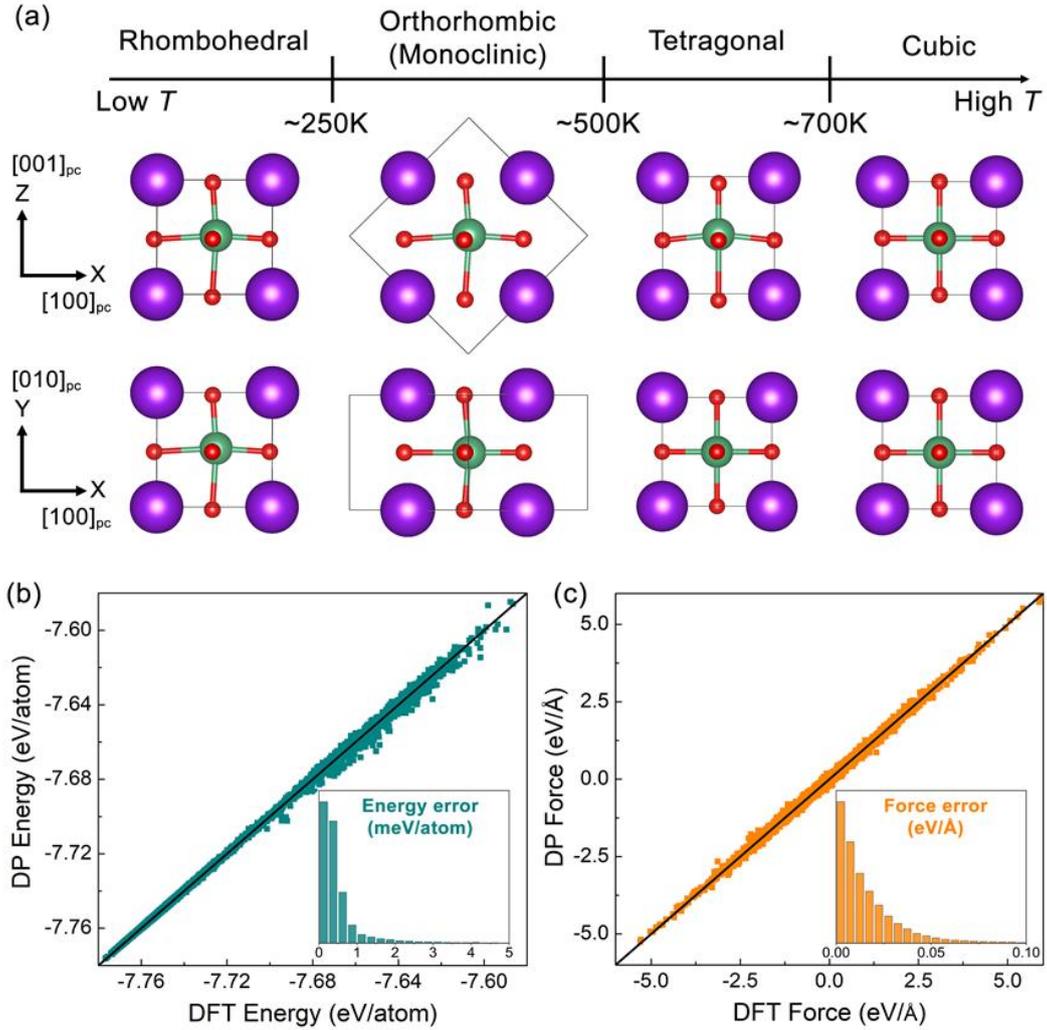

Figure 1. (a) Temperature-dependent phase transition of $KNbO_3$. The purple, green, and red atoms correspond to K, Nb, and O. The 10-atoms orthorhombic cell possesses a 5-atoms monoclinic primitive cell. Comparison of (b) energies and (c) atomic forces between the prediction of DP model and reference DFT calculations. The insets show the distributions of absolute errors.

The prediction of the DP model on the phase stability of $KNbO_3$ is validated by the calculation of the equation of states (EOS) as a function of volume, as shown in Figure 2(a). From the DFT calculation, the energies of the phases at equilibrium state are found following a decreasing sequence, *i.e.*, cubic, tetragonal, orthorhombic, and rhombohedral, which is consistent with the sequence of temperature-dependent phase transition. It is worth mentioning that the EOS curves of the orthorhombic phase and rhombohedral phase are very close to each other, with a very tiny difference

~$10^{-4}$ eV/atom. The EOS curves calculated from DP and DFT are almost identical, indicating that DP can reproduce the sequence of phase stability as effectively as DFT, albeit the tiny difference between orthorhombic and rhombohedral phases.

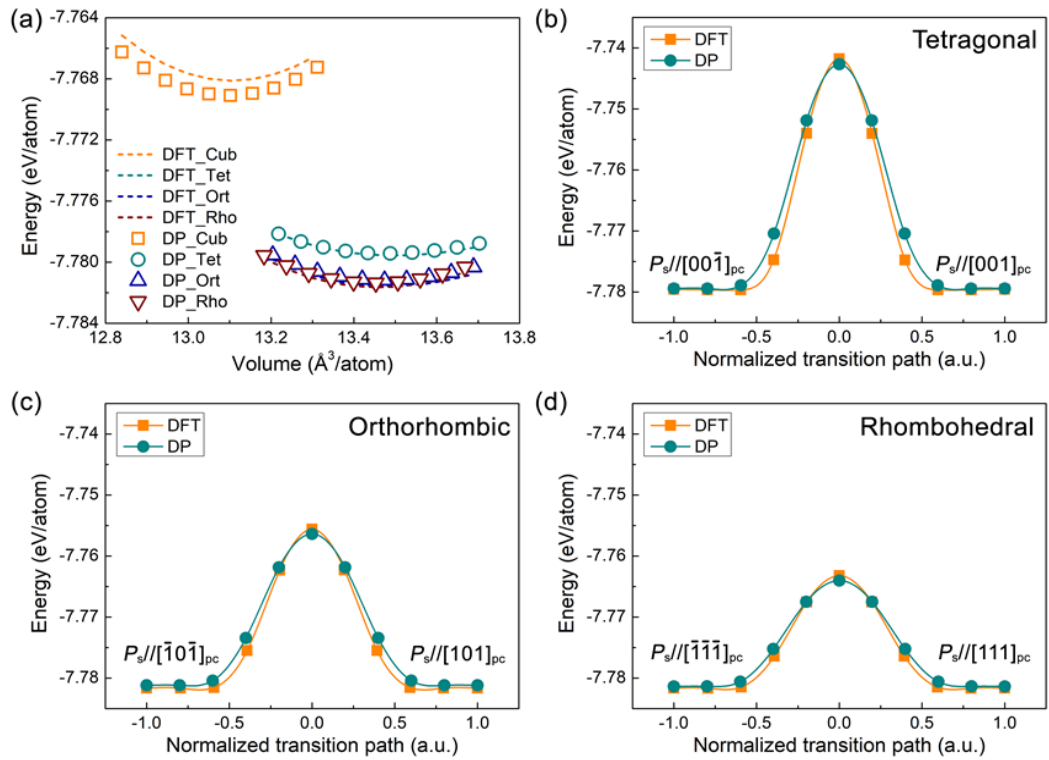

Figure 2. (a) Equation of states of polymorphic $KNbO_3$ as a function of volume. Intrinsic polarization switching in (b) tetragonal phase, (c) orthorhombic phase, and (d) rhombohedral phase.

The switching behavior of spontaneous polarization is one of the most important characteristics of ferroelectrics. Measurement of polarization loop, *i.e.*, switching of polarization, as a function of the electric field is often required to verify the ferroelectricity of a particular material of interest. We simulate the intrinsic polarization switching (opposite to the polarization switching via domain-wall motion) of ferroelectric phases by using the climbing-image nudged elastic band (c-NEB) method in both DP and DFT calculations.[28] For each phase, configurations with spontaneous polarization of antiparallel directions are defined as the initial and final images, so the c-NEB method can determine a minimum-energy transition path between these images. The transition paths obtained in DP and DFT calculations are found nearly overlapped. Minimum energies at initial and final images confirm the favorable configurations exhibiting spontaneous polarization, while the local maxima observed in the transition paths of all ferroelectric phases correspond to the energy barrier of intrinsic polarization switching.

Next, we examine the capability of the DP model in the calculation of elastic properties, as shown in Figure 3. Calculations of elastic properties in DP and DFT were both performed by using the energy-strain method within the finite difference framework, where the deformed configurations are generated with the aids of the VASPKIT package[29] and auto-test module in DP-GEN. Since the number of independent stiffness constants is crystal-symmetry dependent, there are 3, 6, 8, and 6 stiffness constants for cubic, tetragonal, orthorhombic, and rhombohedral phases, respectively. It is observed that the DP and DFT results are very similar, where the mean error is only ~4 GPa. It is noted that these particular deformed configurations have never been added to the training dataset.

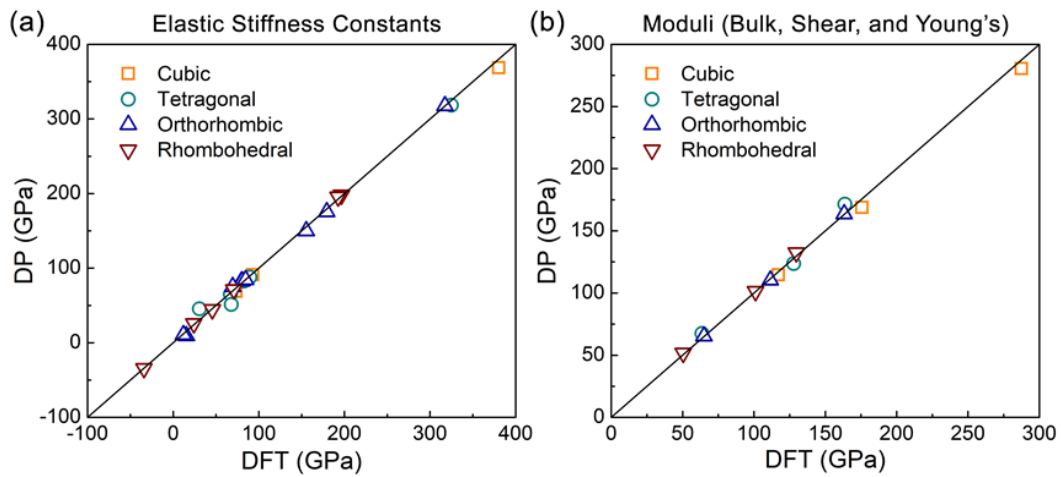

Figure 3. (a) Independent elastic stiffness constants and (b) elastic moduli of cubic, tetragonal, orthorhombic, and rhombohedral phases. Detailed values can be found in Table S1-S5.

Phonon dispersion spectrum is a quantum mechanical description of lattice dynamics along different wave vectors, where the interaction among atoms plays a major role. The comparison of phonon dispersion spectra between DP and DFT calculations is presented in Figure 4. The calculation of phonon dispersion spectra was performed within the finite difference framework by using Phonopy.[30] From Figure 4, it is observed that DP and DFT results are highly alike. The good agreement suggests that the DP model can accurately predict the chemical bonding among atoms in all phases. From the phonon spectra, it is observed that zone-centered soft (or imaginary) phonon mode (with negative frequency) occurs in cubic and tetragonal phases but is absent in orthorhombic and rhombohedral phases. The presence of soft phonon mode indicates the dynamical instability of cubic and tetragonal phases. The concept of soft mode has been used to explain the origin of ferroelectricity and paraelectric-ferroelectric phase transition since the 1960s,[31] where the frequency of soft mode substantially decreases when approaching Curie temperature. It is worth noting that the soft mode does not disappear in the ferroelectric tetragonal phase. It has been noted that a complete softening to zero at paraelectric-ferroelectric phase transition is not necessary.[32]

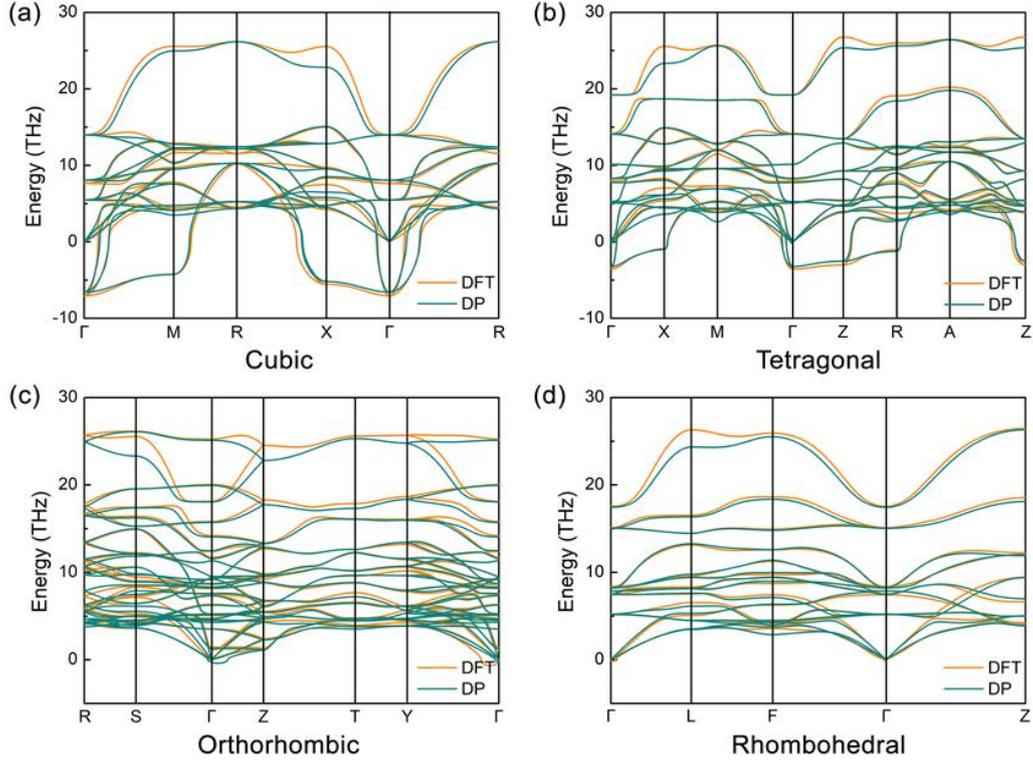

Figure 4. Phonon dispersion spectra of (a) cubic, (b) tetragonal, (c) orthorhombic, and (d) rhombohedral phases.

**B. Domain wall and intrinsic domain-wall motion**

Simulation of domain walls is attempted to verify the transferability of the DP model. As a quick attempt, a 180° domain wall (abbreviated as 180DW) and a 90° domain wall (abbreviated as 90DW) in the orthorhombic $KNbO_3$ were simulated, as shown in Figure 5. Just to mention we have conducted a detailed DFT calculation of orthorhombic $KNbO_3$ in another work, from which a detailed discussion can be found.[33] In the present calculation, the 180DW supercell is composed of 10×1×1 10-atoms orthorhombic unit cells while the 90DW supercell is composed of 10×1×1 5-atoms monoclinic sub-cells. Examples of 180DW and 90DW are shown in Figure 5(a), where the left and right domains are separated by domain walls located in the middle and both ends due to the periodicity. The calculated formation energies of

180DW and 90DW are shown in Figure 5(b). The formation energies of domain walls calculated from DFT and DP show good agreement. The formation energy of 180DW is nearly 40~50 times higher than that of 90DW. Our previous DFT study showed that the Nb-O bonds located at the domain wall region in 180DW drastically change but those in 90DW merely change.[33] Therefore, the huge difference of formation energies between 180DW and 90DW was proposed to be associated with the change of chemical bonding at the domain wall region.

Next, the intrinsic domain-wall motion (*i.e.*, no interference with defects) is investigated by using the c-NEB method, as shown in Figure 5(c-f). The initial and final images correspond to the configurations at states 0 and 1, respectively, where the domain wall located at the middle of the supercell moves across a unit cell (or sub-cell). From both the c-NEB results obtained by DP and DFT shown in Figure 5(e-f), a flat energy curve is observed in the motion of 180DW, while an energy barrier of ~22 mJ/m$^2$ is observed in the motion of 90DW. The difference between the two energy curves has been explained by the dynamical variation of chemical bonding near the domain wall region.

It is impressive that the DFT simulation of domain walls can be well reproduced by the DP simulation, where neither static domain wall configuration nor dynamical configurations during domain-wall motion have been introduced in the training dataset. The capability of DP on domain wall simulation is highly valuable to the investigation of ferroelectric materials because the width of a ferroelectric domain, in reality, is usually at the order of tens nanometers to hundreds of nanometers, which is beyond the reasonable calculation capability of DFT. Our results show that the DP model is beneficial to the future exploration into domain walls of bigger sizes or in different phases. Nevertheless, if the simulation task extrapolates too far from the model, the addition of specific configurations into training might still be necessary.

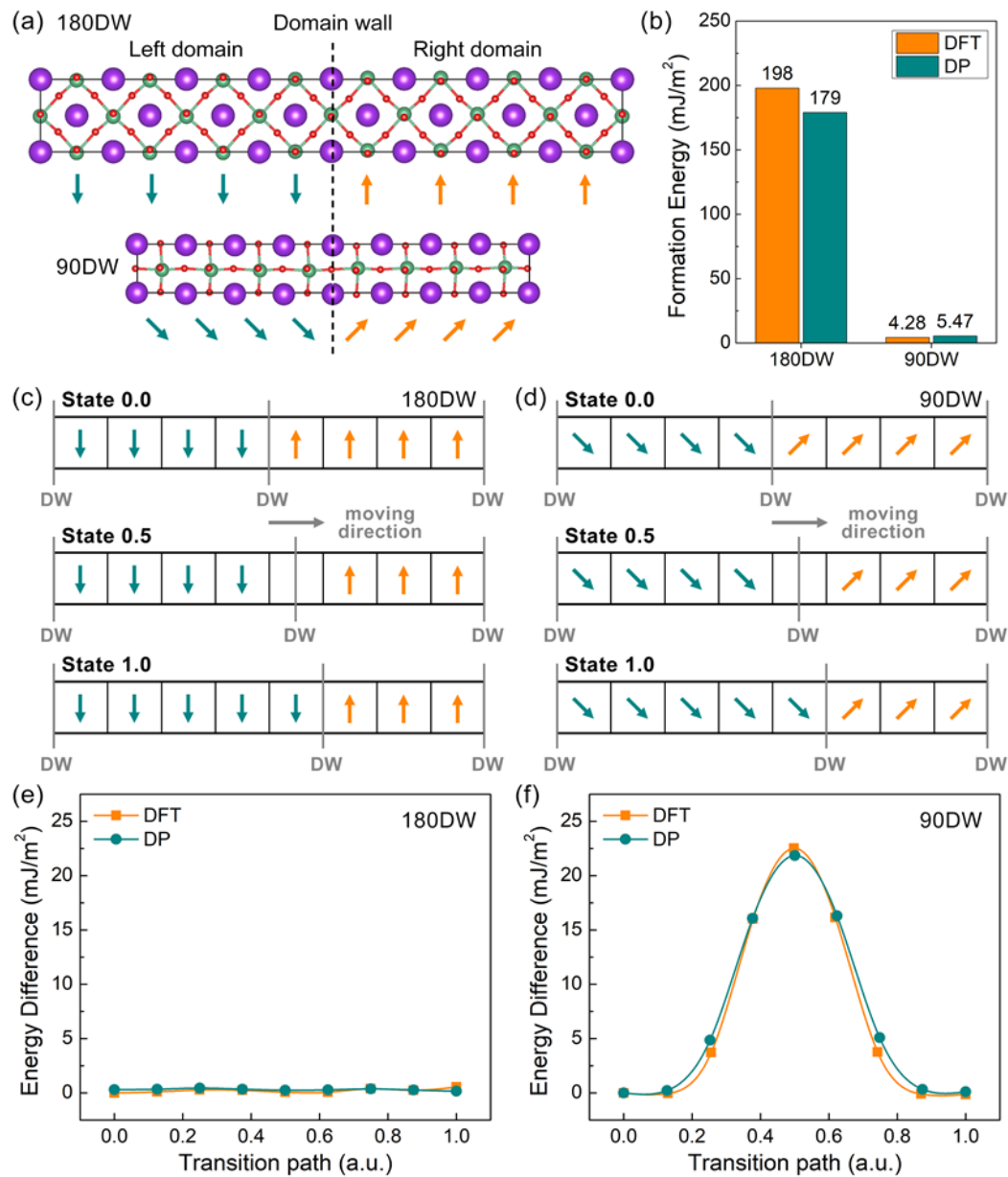

Figure 5. Simulation of 180DW and 90DW in orthorhombic $KNbO_3$. (a) Illustration of domain wall configurations. (b) formation energies of domain walls. Illustration of domain-wall motion of (c) 180DW and (d) 90DW. Calculated transition paths correspond to the domain-wall motions of (e) 180DW and (f) 90DW. Please note that 8×1×1 supercells were illustrated here only for better demonstration, but 10×1×1 supercells were actually calculated.

**C. Phase transition**

Phase transition of $KNbO_3$ is investigated via two approaches, including the variable-cell c-NEB method[28] performed at 0K and temperature-dependent MD simulation. For the variable-cell c-NEB calculation, three different phase transitions were independently calculated and subsequently merged into a continuous transition path as shown in Figure 6, including cubic-tetragonal, tetragonal-orthorhombic, and orthorhombic-rhombohedral phase transitions. A transition path with monotonically decreasing energy is observed during the successive transitions from the cubic phase until the rhombohedral phase. Strangely, no energy barrier appears along the transition path. The simulated continuous evolution of cells in c-NEB calculation is similar to the second-order phase transition. Such simulation is not satisfactory to describe the phase transition of $KNbO_3$, especially the cubic-tetragonal phase transition has been widely known to exhibit strong first-order nature.[34]

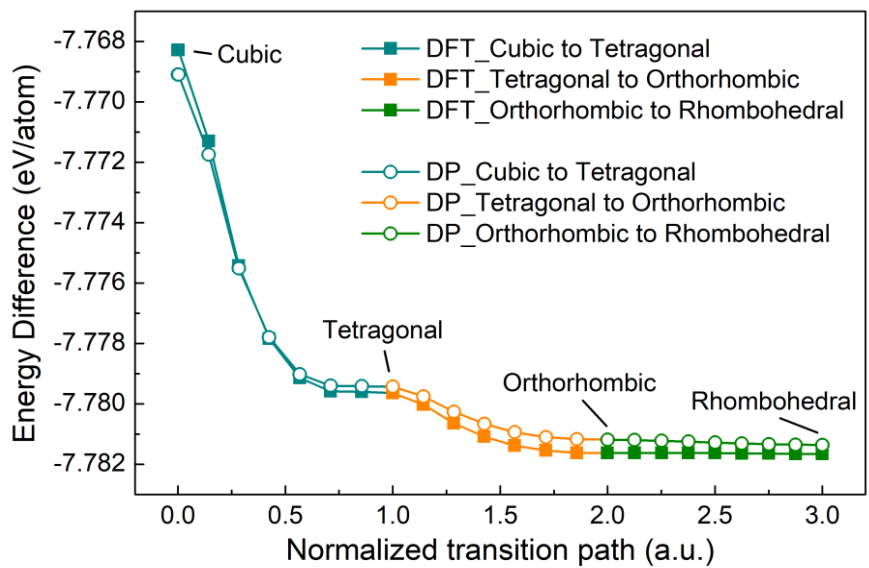

Figure 6. Phase transition in $KNbO_3$ simulated by using variable-cell c-NEB method. Three independent transition paths, which correspond to cubic-tetragonal, tetragonal-orthorhombic, and orthorhombic-rhombohedral phase transitions, were merged and normalized into a continuous transition path.

To better simulate the temperature-dependent phase transition of KNbO$_3$, MD simulation within the NPT ensemble was performed. For the MD simulation, a 13×13×13 supercell composed of 10985 atoms with cubic phase was constructed. The simulation was first initialized at 815 K for 20 ps (with a unit time-step of 1 fs) to ensure stability and later cooled down to 15 K within 160 ps. The simulation result of temperature-dependent phase transition is summarized in Figure 7. Three significant structural transition points can be observed at ~675 K, ~225 K, and ~75 K. Identification of phases is performed by analyzing the "lattice parameters", which are normalized (via averaging) from the simulation box size parameters (13×13×13 supercell). At above ~675 K, lattice parameters of $a≈b≈c$ and $α≈β≈γ≈90°$ correspond to a cubic phase. Within ~675 K to ~225 K, lattice parameters of $a≈b<c$ and $α≈β≈γ≈90°$ correspond to a tetragonal phase. Within ~225 K to ~75 K, lattice parameters of $a≈c>b$ and $α≈β≈γ≈90°$ correspond to an orthorhombic phase (monoclinic as the primitive cell). Finally, at below ~75 K, the lattice feature of an ideal rhombohedral phase ($α=β=γ<90°$) is missing, where a cubic-like lattice ($α≈β≈γ≈90°$) is observed instead. The observation seems to agree with the experimentally known sequence of phase transition upon cooling, but biased shiftings of transition temperatures can be observed. The biased shifting of transition temperatures can be associated with the underestimation/overestimation of lattice parameters, depending on the approximation of the exchange-correlation functional.[10]

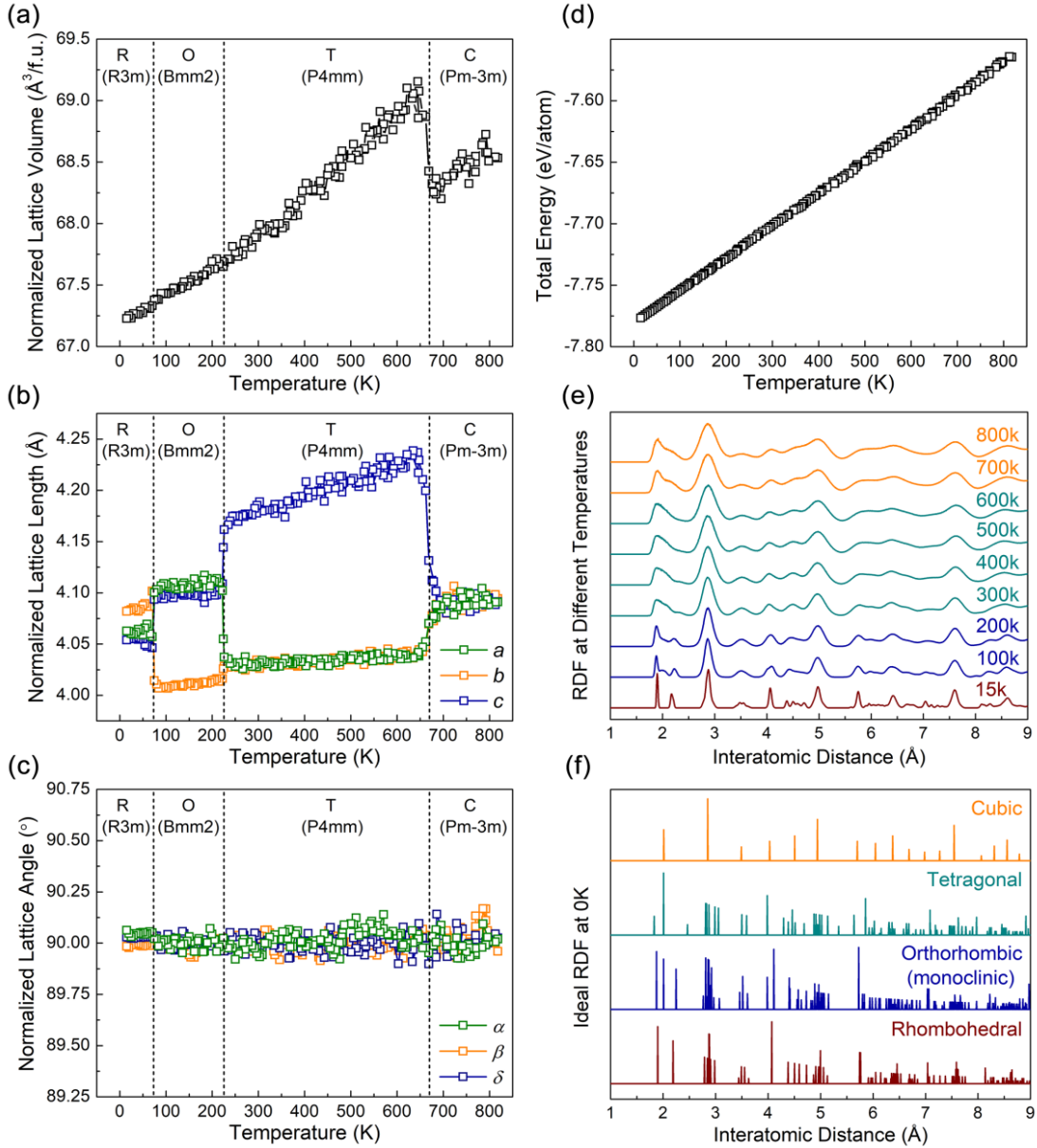

Figure 7. Simulation of temperature-dependent phase transition in KNbO$_3$ within NPT ensemble from 815 K to 15 K. (a) Lattice volume, (b) lattice parameters including *a*, *b*, *c*, and (c) lattice parameters including *α*, *β*, *γ*, normalized (via averaging) from a 13×13×13 supercell. (d) Variation of normalized total energy consists of potential and kinetic energies. (e) Time-averaged RDF at different temperatures. (f) Ideal RDF of 4 major phases at 0K.

To confirm the local lattice structure and atomic configuration, we calculate the time-averaged (every 500 fs) radial distribution function (RDF) of all atoms in the whole simulated box as a function of temperature, as shown in Figure 7e. For comparison, ideal RDFs of four different phases at 0 K are computed as shown in

Figure 7f. It is noted that the RDF at 15 K is highly similar to the ideal RDF of rhombohedral phases, though it shows a diffused distribution due to the thermal fluctuation. The contradiction between the local rhombohedral-like RDF and global cubic-like lattice (observed in Figures 7b and 7c) can be explained by the observed formation of polydomain configuration, as shown in Figure 8. The coexistence of multiple domains with spontaneous polarizations/strains along different directions can eliminate the characteristic lattice parameter feature at the global scale, *e.g.*, the $α=β=γ<90°$ for rhombohedral symmetry. We note that the polydomain configuration does not only occur in the rhombohedral phases but also in the orthorhombic and tetragonal phases. The formation of polydomain can also be easily identified from the analysis of the non-centrosymmetric displacement of Nb atoms among oxygen octahedral, as shown in Figure S2.

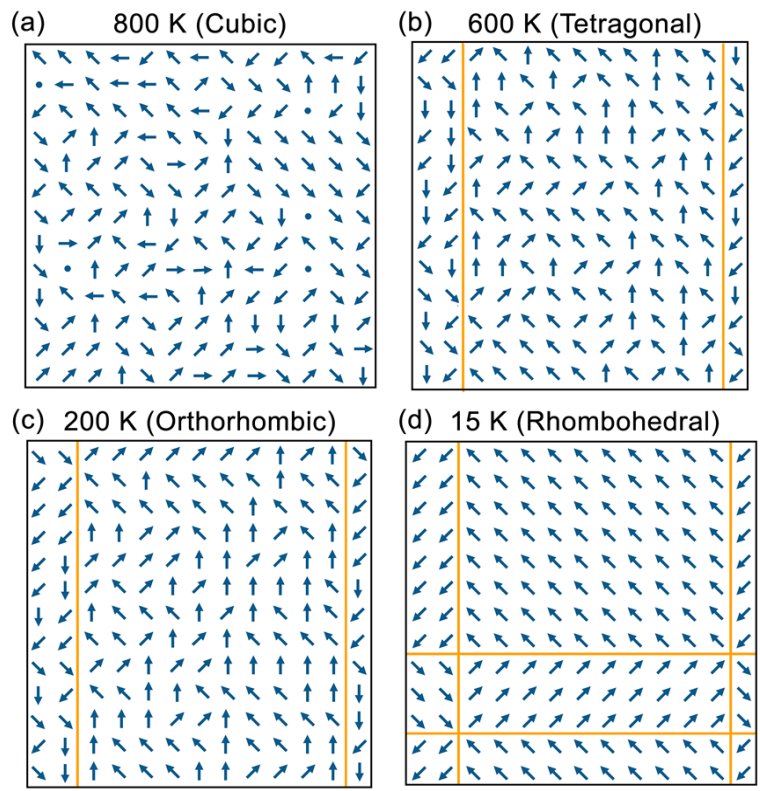

Figure 8. An exaggerated schematic of spontaneous polarization directions in (a) cubic, (b), tetragonal (c), orthorhombic, and (d) rhombohedral phases, where a specific plane of lattices in the simulation supercell is fixed for observation. The estimation of polarization is based on the Nb-O bonding configuration, see Figure S1. Since the estimated polarization directions are projected in 2 dimensions (blue-colored arrow symbols), some polarization directions pointing in/out of the planes cannot be clearly counted (blue-colored dot symbols). Orange-colored lines are provided to show the polydomain configurations.

The formation of domains at the paraelectric-ferroelectric phase transition is usually driven by the minimization of electrostatic energy and strain energy under electromechanical constraints. The formation of polydomain configuration in the present simulation is unexpected since no electromechanical constraint has been defined. As we look at the RDF at 800 K, which corresponds to the paraelectric cubic phase, more than one characteristic Nb-O interatomic pairs have been observed at below 2.5 Å. From Figure S3, the RDF curve at below 2.5 Å can be well fitted by two Gaussian peaks, suggesting a diffused rhombohedral-like RDF (consisting of 3 long bonds and 3 short bonds). The observation seems to agree with the eight-site model,[35] where the disordered paraelectric cubic phase possesses spontaneous polarizations along 8 degenerate $<111>_{pc}$ directions in random order. Upon cooling down to Curie temperature, an "order-disorder" phase transition seems to have occurred. In the tetragonal phase, the possible degenerate direction of spontaneous polarization has become 4, thus a "macroscopic" $<101>_{pc}$ spontaneous polarization can be observed. A similar discussion has been reported in an earlier MD simulation.[36,37] Still, from Figure S2, there is clear evidence showing that some Nb atoms are located at centrosymmetric positions within oxygen octahedron in the paraelectric phase (*e.g.*, 800 K), while most Nb atoms are located at non-centrosymmetric positions within oxygen octahedron in the ferroelectric phase (*e.g.*, 600 K). The observation indicates that the displacive nature is also playing a major role in the phase transition.

## IV. CONCLUSIONS

In summary, a machine-learning interatomic potential of the $KNbO_3$ system was constructed for MD simulation by using the deep learning method. The DP model can accurately reproduce the DFT calculations, including the prediction of atomic force, energy, elastic properties, and phonon dispersion. The DP model was attempted for the simulation of domain wall and temperature-dependent phase transition. The preliminary results show that the DP model is highly potential for these simulation tasks yet more systematic investigation in the future is necessary. We believe the machine-learning interatomic potentials of other ferroelectric perovskites can also be developed via this DP approach. Consequently, theoretical investigation can substantially benefit from it and provide a deeper understanding of ferroelectrics.

## V. ACKNOWLEDGEMENTS

This work was supported by National Nature Science Foundation of China (No. 52032005) and Beijing Natural Science Foundation (No. JQ20009). The authors would like to acknowledge Mr. Jian Han and Dr. Yan-Dong Sun for providing necessary help on the phonon calculation.

**REFERENCES**

[1]     R. W. Whatmore, Y.-M. You, R.-G. Xiong, and C.-B. Eom,  (AIP Publishing LLC, 2021), p. 070401.
[2]     C. Qiu *et al.*, Nature **577**, 350 (2020).
[3]     J. F. Scott and C. A. P. De Araujo, Science **246**, 1400 (1989).
[4]     H. Pan *et al.*, Science **374**, 100 (2021).
[5]     H. Kishi, Y. Mizuno, and H. Chazono, Jpn. J. Appl. Phys. **42**, 1 (2003).
[6]     Z. Li *et al.*, Adv. Funct. Mater. **31**, 2005012 (2021).
[7]     A. J. Bell and O. Deubzer, MRS Bull. **43**, 581 (2018).

[8] T. Zheng, J. Wu, D. Xiao, and J. Zhu, Prog. Mater Sci. **98**, 552 (2018).
[9] H.-C. Thong, A. Payne, J.-W. Li, J. L. Jones, and K. Wang, Acta Mater. **211**, 116833 (2021).
[10] P. Ghosez and J. Junquera, Annu. Rev. Condens. Matter Phys. **13** (2021).
[11] I. Grinberg, Y.-H. Shin, and A. M. Rappe, Phys. Rev. Lett. **103**, 197601 (2009).
[12] X. Zeng and R. Cohen, Appl. Phys. Lett. **99**, 142902 (2011).
[13] M. Sepliarsky, S. Phillpot, M. Stachiotti, and R. Migoni, J. Appl. Phys. **91**, 3165 (2002).
[14] Y.-H. Shin, I. Grinberg, I.-W. Chen, and A. M. Rappe, Nature **449**, 881 (2007).
[15] M. Sepliarsky, Z. Wu, A. Asthagiri, and R. Cohen, Ferroelectrics **301**, 55 (2004).
[16] Y.-H. Shin, V. R. Cooper, I. Grinberg, and A. M. Rappe, Phys. Rev. B **71**, 054104 (2005).
[17] J. C. Wojdeł, P. Hermet, M. P. Ljungberg, P. Ghosez, and J. Iniguez, J. Phys.: Condens. Matter **25**, 305401 (2013).
[18] J. Behler, The Journal of chemical physics **145**, 170901 (2016).
[19] T. Mueller, A. Hernandez, and C. Wang, The Journal of chemical physics **152**, 050902 (2020).
[20] L. Zhang, J. Han, H. Wang, R. Car, and E. Weinan, Phys. Rev. Lett. **120**, 143001 (2018).
[21] F. Noé, S. Olsson, J. Köhler, and H. Wu, Science **365** (2019).
[22] J. Han, L. Zhang, and R. Car, arXiv preprint arXiv:1707.01478 (2017).
[23] C. Wang, J. Wu, Z. Zeng, J. Embs, Y. Pei, J. Ma, and Y. Chen, npj Computational Materials **7**, 1 (2021).
[24] J. Wu, Y. Zhang, L. Zhang, and S. Liu, Phys. Rev. B **103**, 024108 (2021).
[25] Y. Zhang, H. Wang, W. Chen, J. Zeng, L. Zhang, H. Wang, and E. Weinan, Comput. Phys. Commun. **253**, 107206 (2020).
[26] H. Wang, L. Zhang, J. Han, and E. Weinan, Comput. Phys. Commun. **228**, 178 (2018).
[27] M. Fontana, G. Metrat, J. Servoin, and F. Gervais, J. Phys. C Solid State Phys. **17**, 483 (1984).
[28] D. Sheppard, P. Xiao, W. Chemelewski, D. D. Johnson, and G. Henkelman, J. Chem. Phys. **136**, 074103 (2012).
[29] V. Wang, N. Xu, J.-C. Liu, G. Tang, and W.-T. Geng, Comput. Phys. Commun., 108033 (2021).
[30] A. Togo and I. Tanaka, Scripta Mater. **108**, 1 (2015).
[31] W. Cochran, Phys. Rev. Lett. **3**, 412 (1959).
[32] S. Kamba, APL Materials **9**, 020704 (2021).
[33] H.-C. Thong, B. Xu, and K. Wang, Appl. Phys. Lett. **120**, 052902 (2022).
[34] L. E. Cross and G. A. Rossetti Jr, J. Appl. Phys. **69**, 896 (1991).
[35] R. Comes, M. Lambert, and A. Guinier, Solid State Commun. **6**, 715 (1968).
[36] M. Sepliarsky, M. Stachiotti, R. Migoni, and C. Rodriguez, Ferroelectrics **234**, 9 (1999).
[37] Z. Tan, Y. Peng, J. An, Q. Zhang, and J. Zhu, lnorg. Chem. **60**, 7961 (2021).

Supplementary information

# Machine-learning interatomic potential for molecular dynamics simulation of ferroelectric KNbO$_3$ perovskite

Hao-Cheng Thong,[1,2] XiaoYang Wang,[3,*] Han Wang,[3] Linfeng Zhang,[4] Ke Wang[1,*], and Ben Xu[2,*]

**AFFILIATIONS**

[1] State Key Laboratory of New Ceramics and Fine Processing, School of Materials Science and Engineering, Tsinghua University, Beijing 100084, PR China.

[2] Graduate School, China Academy of Engineering Physics, Beijing 100193, PR China.

[3] Laboratory of Computational Physics, Institute of Applied Physics and Computational Mathematics, Huayuan Road 6, Beijing 100088, PR China.

[4] Beijing Institute of Big Data Research, Beijing 100871, PR China.

*Author to whom correspondence should be addressed: xiaoyanglanl@gmail.com (X. Wang), wang-ke@tsinghua.edu.cn (K. Wang), bxu@gscaep.ac.cn (B. Xu)

Table S1. Elastic stiffness tensor of the cubic phase. (unit in GPa)

|     | $C_{11}$ | $C_{12}$ | $C_{44}$ |
|-----|----------|----------|----------|
| DFT | 380.34   | 73.34    | 92.88    |
| DP  | 368.48   | 68.89    | 91.21    |

Table S2. Elastic stiffness tensor of the tetragonal phase. (unit in GPa)

|     | $C_{11}$ | $C_{12}$ | $C_{13}$ | $C_{33}$ | $C_{44}$ | $C_{66}$ |
|-----|----------|----------|----------|----------|----------|----------|
| DFT | 324.90   | 82.67    | 66.85    | 67.97    | 30.67    | 89.59    |
| DP  | 318.41   | 82.75    | 64.74    | 51.29    | 45.57    | 88.04    |

Table S3. Elastic stiffness tensor of the orthorhombic phase. (unit in GPa)

|     | $C_{11}$ | $C_{12}$ | $C_{13}$ | $C_{22}$ | $C_{23}$ | $C_{33}$ | $C_{44}$ | $C_{55}$ | $C_{66}$ |
|-----|----------|----------|----------|----------|----------|----------|----------|----------|----------|
| DFT | 179.34   | 79.65    | 15.38    | 317.43   | 80.39    | 155.49   | 69.43    | 11.97    | 84.61    |
| DP  | 175.52   | 82.41    | 9.30     | 317.66   | 82.82    | 149.94   | 75.16    | 10.71    | 84.41    |

Table S4. Elastic stiffness tensor of the rhombohedral phase. (unit in GPa)

|     | $C_{11}$ | $C_{12}$ | $C_{13}$ | $C_{14}$ | $C_{33}$ | $C_{44}$ |
|-----|----------|----------|----------|----------|----------|----------|
| DFT | 196.54   | 70.69    | 45.78    | -34.20   | 192.65   | 24.02    |
| DP  | 197.86   | 71.16    | 44.40    | -34.53   | 195.43   | 25.52    |

Table S5. Elastic moduli and Poisson's ratio of different phases. (unit in GPa)

|                  | Cubic  |        | Tetragonal |        | Orthorhombic |        | Rhombohedral |        |
|------------------|--------|--------|------------|--------|--------------|--------|--------------|--------|
|                  | DP     | DFT    | DP         | DFT    | DP           | DFT    | DP           | DFT    |
| Bulk modulus     | 168.75 | 175.68 | 123.71     | 127.83 | 110.29       | 111.45 | 101.23       | 101.14 |
| Shear modulus    | 114.64 | 117.13 | 67.56      | 63.61  | 65.29        | 64.99  | 51.62        | 50.42  |
| Youngs modulus   | 280.42 | 287.49 | 171.46     | 163.68 | 163.60       | 163.24 | 132.37       | 129.72 |
| Poisson ratio    | 0.22   | 0.23   | 0.27       | 0.29   | 0.25         | 0.26   | 0.28         | 0.29   |

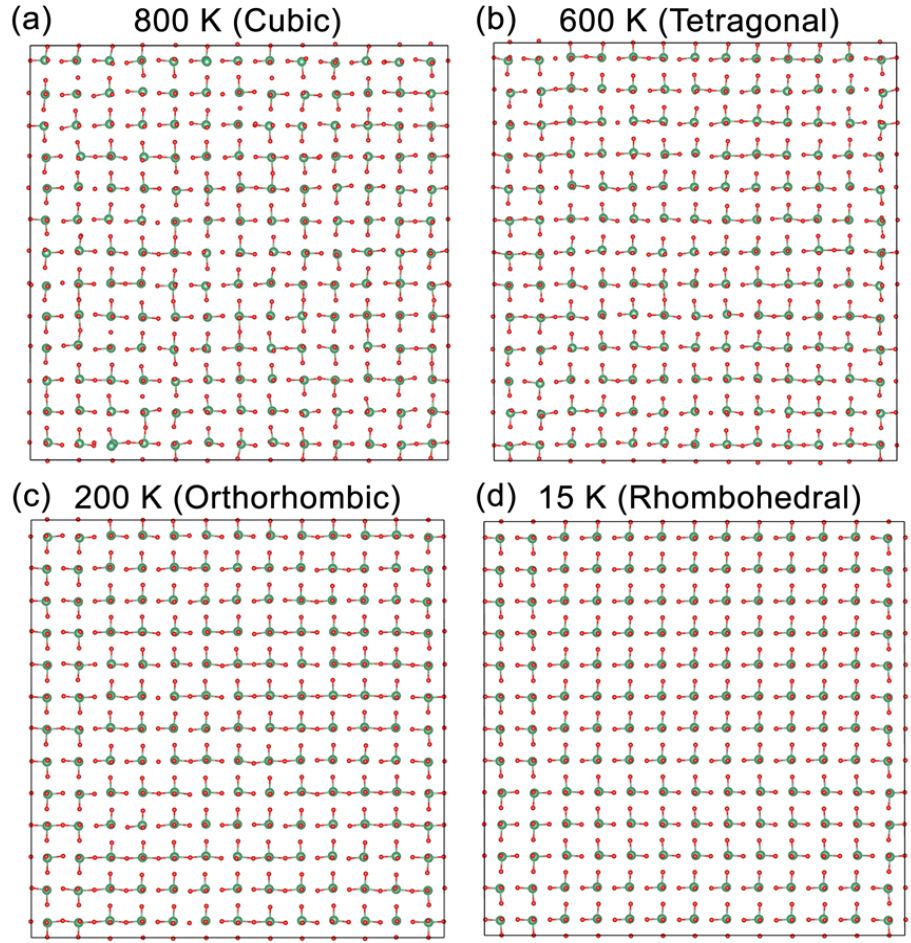

Figure S1. Nb-O bonding configuration in (a) cubic, (b), tetragonal (c), orthorhombic, and (d) rhombohedral phases, where a specific plane of lattices in the simulation supercell is fixed for observation. The green and red atoms represent Nb and O, respectively. A bond is sketched if the interatomic distance is smaller than 2.05 Å.

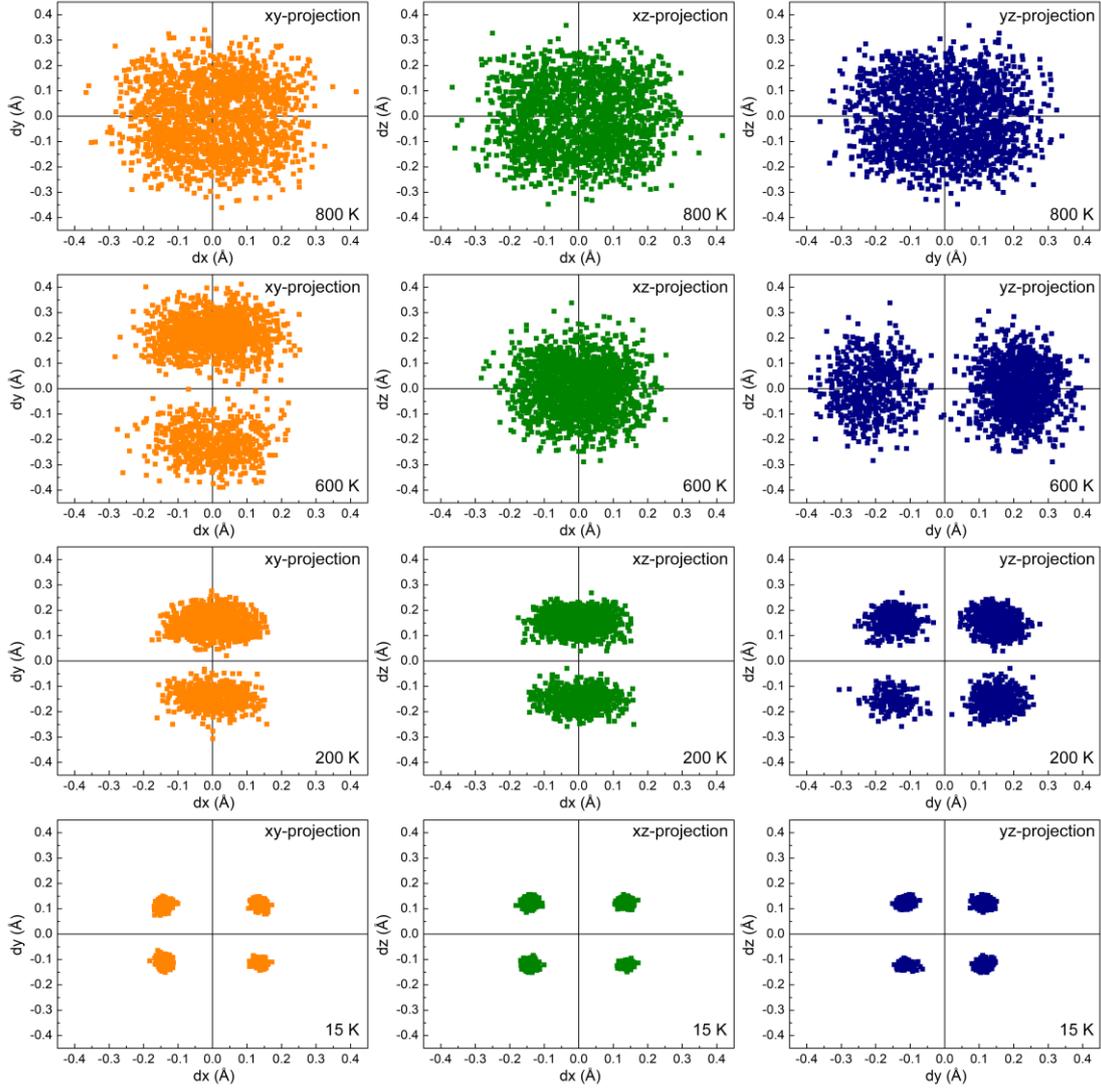

Figure S2. Relative displacements of Nb atoms from the center of the oxygen octahedron calculated from the temperature-dependent MD simulation. The first, second, third, and fourth rows correspond to the data obtained at 800 K (cubic), 600 K (tetragonal), 200 K (orthorhombic), and 15 K (rhombohedral).

The relative displacement vector $\vec{d}$ is calculated as:

$$\vec{d} = \vec{Nb} - (\vec{O_1} + \vec{O_2} + \vec{O_3} + \vec{O_4} + \vec{O_5} + \vec{O_6})/6,$$

where $\vec{Nb}$ is a vector pointing from the origin to the coordinate of the Nb atom, while $\vec{O_x}$ ($x$=1, 2, 3, 4, 5, 6) are vectors pointing from the origin to the six nearest O atoms around the Nb atom.

Judging from the relative displacement of Nb atoms from the center of the oxygen octahedron, one may roughly estimate the polarization direction. At 600 K (tetragonal), domains with polarization directions along $x^0y^+z^0$ and $x^0y^-z^0$ can be observed. At 200 K (orthorhombic), domains with polarization directions along $x^0y^+z^+$, $x^0y^+z^-$, $x^0y^-z^+$, and $x^0y^-z^-$ can be observed. At 15 K (rhombohedral), domains with polarization directions along $x^+y^+z^+$, $x^+y^+z^-$, $x^+y^-z^+$, $x^+y^-z^-$, $x^-y^+z^+$, $x^-y^+z^-$, $x^-y^-z^+$, $x^-y^-z^-$ can be observed. The superscript $^0$, $^+$, and $^-$ correspond to the zero polarization, polarization parallel with the axis, and polarization antiparallel with the axis, respectively.

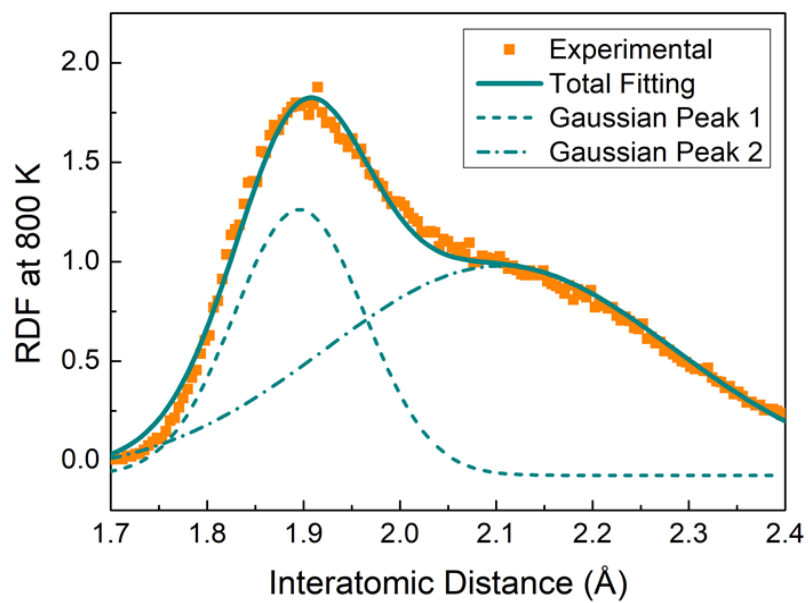

Figure S3. The fitting of RDF at 800 K. Two Gaussian peaks are used for the fitting assuming that there are Nb-O long bonds and short bonds. The $R^2$ of the fitting is 0.989, showing the fitting is reasonable. The center of peak 1 and peak 2 are 1.896 Å and 2.104 respectively.